\documentclass[preprint]{aastex}

 \usepackage{emulateapj5}
 \usepackage{apjfonts}

\newcommand{\Dnu}[1]{\Delta \nu_{#1}}
\newcommand{\dnu}[1]{\delta \nu_{#1}}
\newcommand{\acena}{\mbox{$\alpha$~Cen~A}}
\newcommand{\acenb}{\mbox{$\alpha$~Cen~B}}
\newcommand{\acen}{\mbox{$\alpha$~Cen}}
\newcommand{\bhyi}{\mbox{$\beta$~Hyi}}
\newcommand{\dpav}{\mbox{$\delta$~Pav}}
\newcommand{\cms}{\mbox{cm\,s$^{-1}$}}

\newcommand{\ms}{\mbox{m\,s$^{-1}$}}
\newcommand{\muHz}{\mbox{$\mu$Hz}}

\newcommand{\new}[1]{\relax #1}
\newcommand{\newnew}[1]{\relax #1}
\newcommand{\half}{{\textstyle\frac{1}{2}}}
\newcommand{\sixth}{{\textstyle\frac{1}{6}}}
\newcommand{\tenth}{{\textstyle\frac{1}{10}}}

\slugcomment{Submitted to ApJ}

\shorttitle{Oscillations in $\alpha$ Cen B}
\shortauthors{Kjeldsen et al.}

\begin{document}

\title{Solar-like oscillations in $\alpha$~Centauri~B}

\author{
Hans Kjeldsen,\altaffilmark{1}
Timothy R. Bedding,\altaffilmark{2}
R. Paul Butler,\altaffilmark{3}
J{\o}rgen Christensen-Dalsgaard,\altaffilmark{1}
Laszlo L. Kiss,\altaffilmark{2}
Chris McCarthy,\altaffilmark{3}
Geoffrey W. Marcy,\altaffilmark{4} 
Christopher G. Tinney\altaffilmark{5} and
Jason T. Wright\altaffilmark{4}
}

\altaffiltext{1}{Department of Physics and Astronomy, University of Aarhus,
DK-8000 Aarhus C, Denmark; hans@phys.au.dk, jcd@phys.au.dk}

\altaffiltext{2}{School of Physics A28, University of Sydney, NSW 2006,
Australia; bedding@physics.usyd.edu.au, laszlo@physics.usyd.edu.au}

\altaffiltext{3}{Carnegie Institution of Washington,
Department of Terrestrial Magnetism, 5241 Broad Branch Road NW, Washington,
DC 20015-1305; paul@dtm.ciw.edu, chris@dtm.ciw.edu}

\altaffiltext{4}{Department of Astronomy, University of California,
Berkeley, CA 94720; and Department of Physics and Astronomy, San Francisco,
CA 94132; gmarcy@astron.berkeley.edu}

\altaffiltext{5}{Anglo-Australian Observatory, P.O.\,Box 296, Epping, NSW
1710, Australia; cgt@aaoepp.aao.gov.au}

\begin{abstract} 
We have made velocity observations of the star \acenb{} from two sites,
allowing us to identify 37 oscillation modes with $l=0$--$3$.  Fitting to
these modes gives the large and small frequency separations as a function
of frequency.  The mode lifetime, as measured from the scatter of the
oscillation frequencies about a smooth trend, is similar to that in the
Sun.  Limited observations of the star \dpav{} show oscillations centred at
2.3\,mHz with peak amplitudes close to solar.  We introduce a new method of
measuring oscillation amplitudes from heavily-smoothed power density
spectra, from which we estimated amplitudes for \acena~and~B, \bhyi,
\dpav{} and the Sun.  We point out that the oscillation amplitudes may
depend on which spectral lines are used for the velocity measurements.
\end{abstract}

\keywords{stars: individual (\acena, \acenb, \bhyi, \dpav) ---
stars:~oscillations --- Sun:~helioseismology}

\section{Introduction}

The $\alpha$~Cen A \& B system is an excellent target for asteroseismology.
Velocity oscillations in the A component were measured by
\citet{B+C2001,B+C2002} and subsequently by \citet[][see also
\citealt{BKB2004}]{BBK2004}.  The first detection of oscillations in
\acenb{} (HR~5460) was made by \citet{C+B2003}, based on velocity
measurements with the CORALIE spectrograph in Chile spanning 13 nights.
They observed excess power centred at 4\,mHz and used the autocorrelation
of the power spectrum to measure a large separation of 161.1\,\muHz.  They
identified twelve oscillation frequencies with $l = 0$--$2$, although the
high sidelobes from their single-site observations, coupled with the
relatively low signal-to-noise (2.5--3.5), means that these may have been
affected by daily aliases.  Theoretical modelling of \acena{} and~B has
been carried out several times, most recently by \citet{ECT2004}, who also
give a thorough review of previous work.

Here we report observations of \acenb{} made from two sites that show
oscillations clearly and allow us to measure their frequencies, amplitudes
and mode lifetimes.

\section{Velocity observations and power spectra}

We observed \acenb{} in May 2003.  At the European Southern Observatory in
Chile we used UVES (UV-Visual Echelle Spectrograph) at the 8.2-m Unit
Telescope~2 of the Very Large Telescope (VLT)\footnote{Based on
observations collected at the European Southern Observatory, Paranal, Chile
(ESO Programme 71.D-0618)}\@.  At Siding Spring Observatory in Australia we
used UCLES (University College London Echelle Spectrograph) at the 3.9-m
Anglo-Australian Telescope (AAT).  In both cases, an iodine cell was used
to provide a stable wavelength reference \citep{BMW96}.

At the VLT we obtained 3379 spectra of \acenb, with typical exposure times
of 4\,s and a median cadence of one exposure every 32\,s.  At the AAT we
obtained 1642 spectra, with typical exposure times of 10--12\,s (but
sometimes as long as 30\,s in poor weather) and a median cadence of one
exposure every 91\,s.  Note that, unlike for our observations of \acena,
UCLES was used in standard planet-search mode.  For \acena{} we rotated the
CCD by 90 degrees to speed up readout time, but found that this introduced
drifts and sudden jumps in the velocities that we believe to be related to
the movement of the CCD dewar as liquid nitrogen boiled off and was
refilled \citep{BBK2004}.

\begin{figure*}
\epsscale{0.7}
\plotone{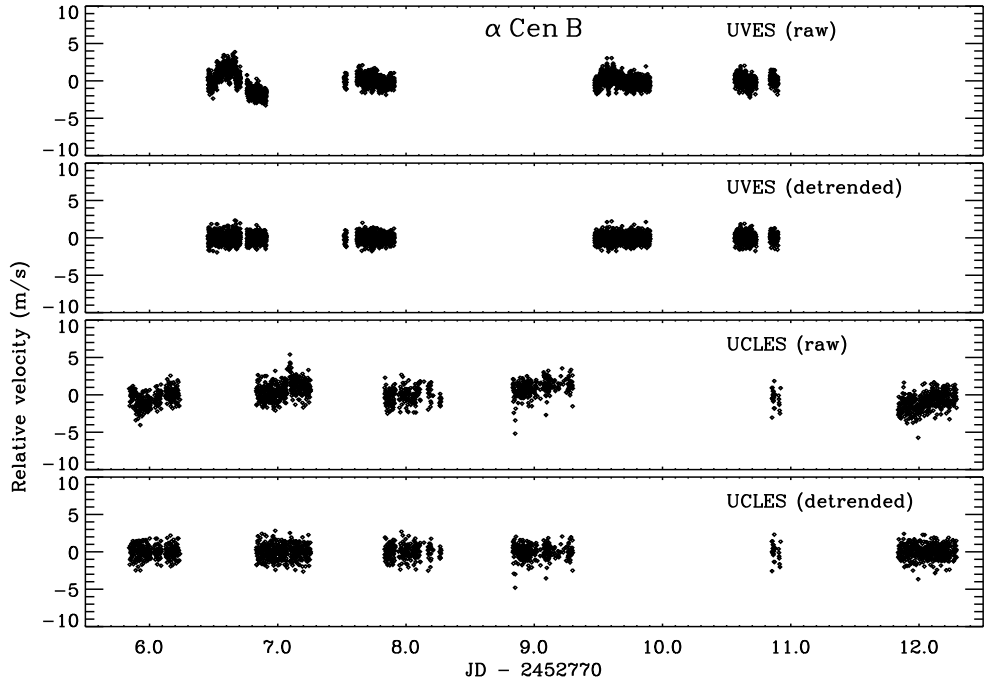}
\caption[]{\label{fig.series} Time series of velocity measurements of
\acenb, both before and after removal of slow trends.  }
\end{figure*}

The resulting velocities are shown in Fig.~\ref{fig.series},
and the effects of bad weather can be seen (we were allocated five nights
with the VLT and eight with the AAT).  Both sets of velocities show trends
during each night that are presumably due to a combination of instrumental
drift and stellar convection noise.  To remove these trends, we have
subtracted a smoothed version of the data one night at a time, producing
the detrended time series shown in Fig.~\ref{fig.series}.  This process of
high-pass filtering is not strictly necessary, but it does slightly reduce
the noise leaking into higher frequencies that would otherwise degrade the
oscillation spectrum.

\new{Note that Fourier analysis on unevenly spaced data cannot use the Fast
Fourier Transform (FFT) algorithm.  We instead employed the method used for
many years by various groups, including our own: we calculated the discrete
Fourier transform, but with individual data points being weighted according
to their quality \citep[e.g.][]{Dee75,FJK95}.}

\begin{figure*}
\epsscale{0.7}
\plotone{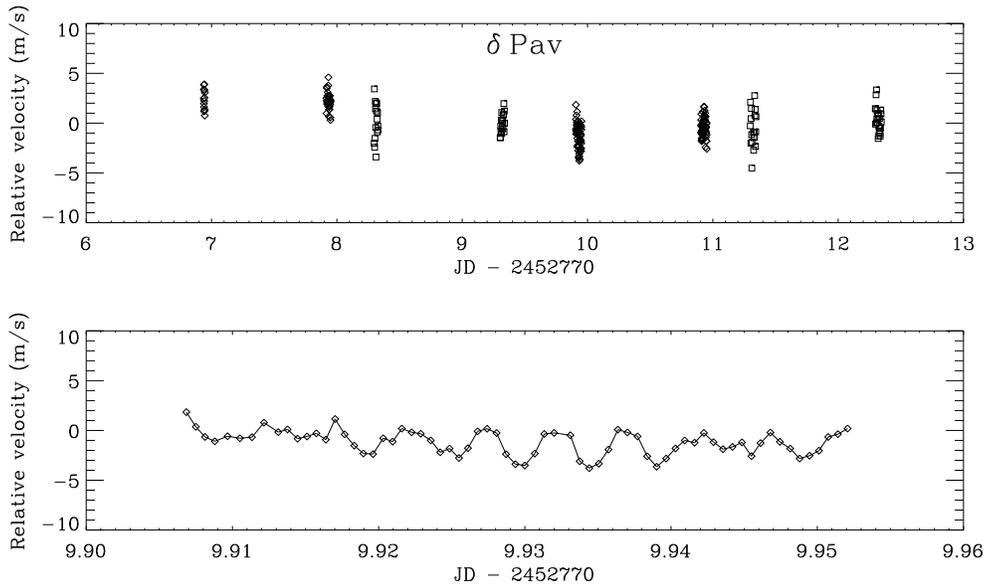}
\caption[]{\label{fig.dpav.series} Time series of velocity measurements of
\dpav{} from UVES (diamonds) and UCLES (squares). The lower panel shows a
1.1-hr segment of UVES velocities, in which the 7-min periodicity is
clearly visible.  }
\end{figure*}

\begin{figure*}
\epsscale{0.7}
\plotone{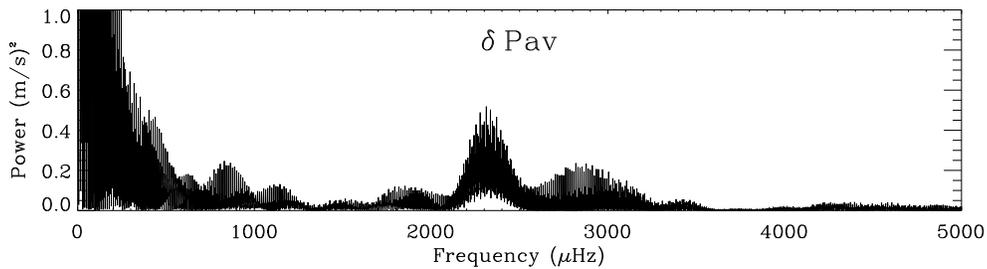}
\caption[]{\label{fig.dpav.power} Power spectrum of velocity measurements of
\dpav{} from UVES and UCLES\@.  There is a clear power excess at 2--3\,mHz.  }
\end{figure*}

For about one hour at the end of each night, when \acenb{} was
inaccessible, we observed the star \dpav{} (HR\,7665; HIP\,99240; G8\,V).  We
obtained a total of 179 spectra with the VLT (median cadence 58\,s) and 77
with the AAT (median cadence 181\,s).  The velocities are shown in
Fig.~\ref{fig.dpav.series} and there is a clear periodicity of about
7\,min.  We did not expect to be able to measure frequencies with such a
limited data set, but we do see a clear excess in the power spectrum from
the oscillations (Fig.~\ref{fig.dpav.power}), whose amplitudes we discuss
in Sec.~\ref{sec.amp}.

Our analysis of the velocities for \acenb{} follows the method that we
developed for \acena{} \citep{BBK2004}.  We have used the measurement
uncertainties, $\sigma_i$, as weights in calculating the power spectrum
(according to $w_i = 1/\sigma_i^2$), but modified some of the weights to
account for a small fraction of bad data points.  In this case, only three
data points from UVES and none from UCLES needed to be down-weighted.
After these adjustments, we measured the average noise in the amplitude
spectrum at high frequencies (above the stellar signal) to be 1.49\,\cms{}
for UVES (6--8\,mHz) and 4.28\,\cms{} for UCLES (4.75--5.50\,mHz).  Note
that we used a lower frequency range to measure the noise in the UCLES data
because the sampling time of those observations was significantly longer
than for UVES (see above).

\begin{figure}
\epsscale{0.8}
\plotone{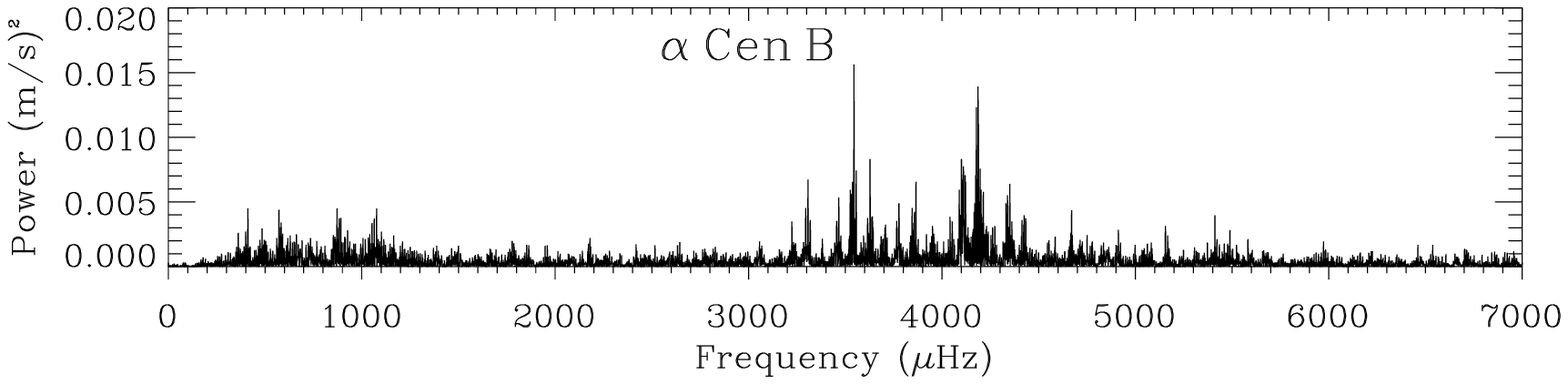}
\medskip
\plotone{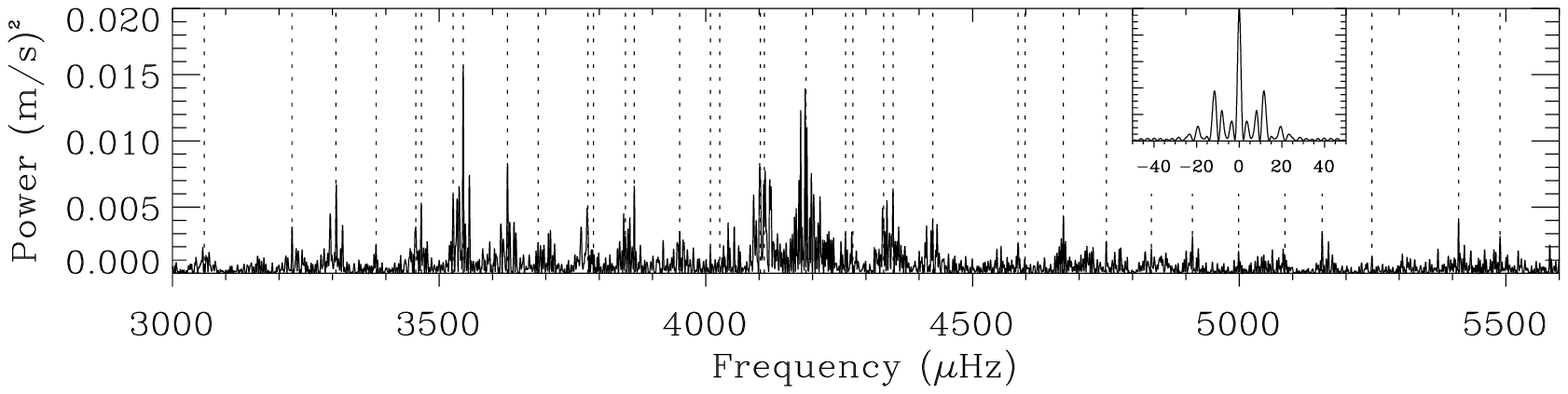}
\medskip
\plotone{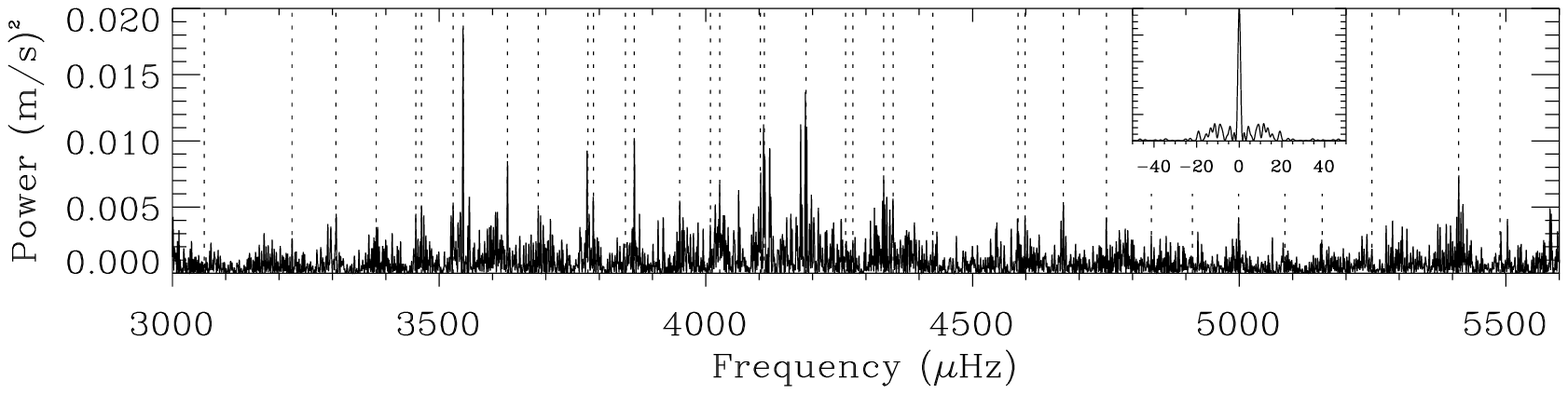}
\caption[]{\label{fig.power} Power spectrum of \acenb{} from the combined
  UVES and UCLES data.  Top panel: the noise-optimized power spectrum, in
  which weights were based on the measurement uncertainties, which
  minimizes the noise.  Middle panel: same as top panel, but expanded to
  show only the central region.  Bottom panel: the sidelobe-optimized power
  spectrum, in which weights were adjusted to minimize the aliases.  The
  spectral windows are shown as insets and vertical dotted lines show our
  final oscillation frequencies, as listed in Table~\ref{tab.freqs}. }
\end{figure}

The power spectrum of the combined series is shown in the top panel of
Fig.~\ref{fig.power}.  We refer to this as the noise-optimized power
spectrum because the weights have been chosen to minimize the noise.  The
noise level in amplitude is 1.39\,\cms{} (6--8\,mHz).  As such, these
measurements replace our observations of \acena{} (1.9\,\cms) as the most
precise amplitude spectrum obtained on any star apart from the Sun.

The middle panel of Fig.~\ref{fig.power} shows a close-up of the
noise-optimized power spectrum.  There is a clear series of regular peaks
with a spacing of about 81\,\muHz, and we therefore confirm that the large
separation is about 162\,\muHz, in agreement with the value reported by
\citet{C+B2003}.  The inset shows the spectral window, in which we see
prominent sidelobes (38\% in power) due to gaps in the observing window.

As for \acena{}, we have also generated a power spectrum in which the
weights were adjusted on a night-by-night basis in order to to minimize the
sidelobes.  The result, which we refer to as the sidelobe-optimized power
spectrum, is shown in the bottom panel of the figure.  The values for these
weight multipliers were: 1.0 for all UCLES nights except the last, which
had 0.8; and 3.2, 3.2, 2.3 and 2.3 for the four UVES nights.  Adjusting the
weights in this way increased the noise level to 2.40\,\cms{} (6--8\,mHz),
but allowed us to identify correctly peaks that might otherwise be obscured
by sidelobes from their stronger neighbours (now reduced to 13\% in power).
In addition, the sidelobe minimization has slightly improved the frequency
resolution, with the FWHM of the spectral window decreasing by about 20\%
(from 1.83\,\muHz{} to 1.44\,\muHz).  This is because of the increased
weight given to the UCLES data, which covers the longer timespan.  Note
that the dotted lines in Fig.~\ref{fig.power} show our final oscillation
frequencies, which are discussed in detail in the next section.

\section{Oscillation frequencies}  \label{sec.freqs}

\begin{figure}
\epsscale{0.5}
\plotone{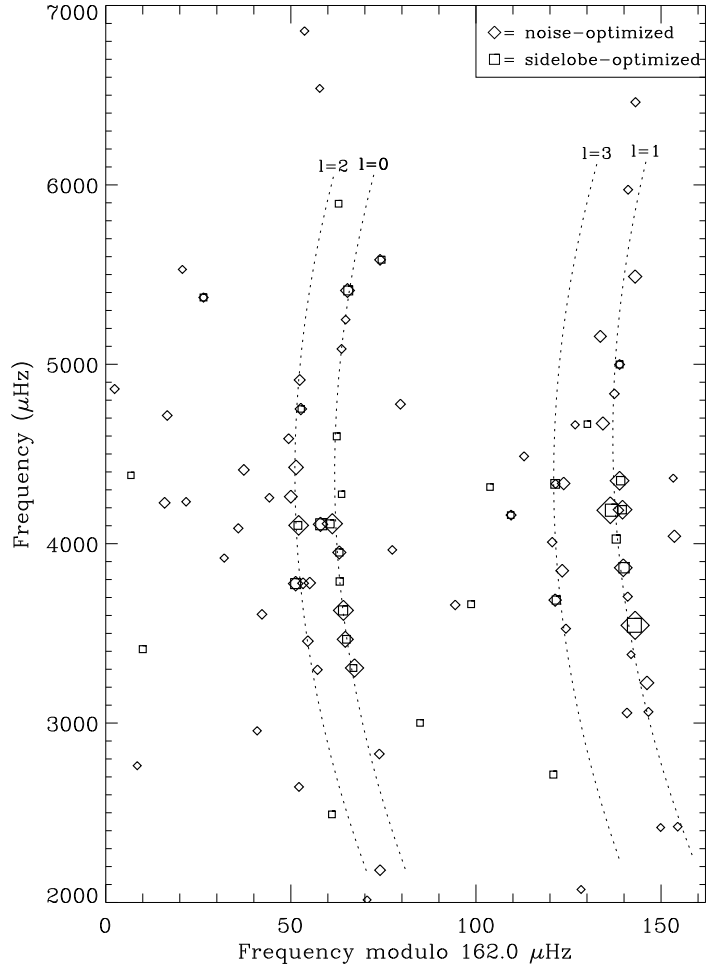}
\caption[]{\label{fig.echelle} Peaks extracted by iterative sine-wave
  fitting to the \acenb{} power spectra, displayed in echelle format.
  Diamonds and squares, respectively, show peaks extracted from the
  noise-optimized and sidelobe-optimized power spectra.  Symbol sizes are
  proportional to the SNR of the peaks.  The dotted curves are fits to the
  final frequencies, given by equations~(\ref{eq.nu0})--(\ref{eq.nu3}).  }
\end{figure}

We measured the frequencies of the strongest peaks in the power spectrum in
the standard way, using iterative sine-wave fitting.  We did this for both
the noise-optimized and the sidelobe-optimized power spectra, and the
resulting frequencies are plotted in echelle format in
Fig.~\ref{fig.echelle}.  The echelle format takes advantage of the fact
that mode frequencies for low-degree p-mode oscillations are approximated
reasonably well by the asymptotic relation:
\begin{equation}
  \nu_{n,l} = \Dnu{} (n + \half l + \epsilon) - l(l+1) D_0.
        \label{eq.asymptotic}
\end{equation}
Here $n$ (the radial order) and $l$ (the angular degree) are integers,
$\Dnu{}$ (the large separation) reflects the average stellar density, $D_0$
is sensitive to the sound speed near the core and $\epsilon$ is sensitive
to the surface layers.

Symbol size in Fig.~\ref{fig.echelle} indicates the signal-to-noise ratio
(SNR) of the peaks, and all peaks with SNR $\ge 2.5$ are shown.  Many of
these represent oscillations but some are artefacts due to the non-linear
nature of the iterative fitting method and its interaction with noise.  The
dotted lines represent a fit to the final frequencies that is discussed
below; in this figure, they serve to guide the eye and allow us to identify
ridges with angular degrees of $l=0$, 1, 2 and~3.

We can see from Fig.~\ref{fig.echelle} that many peaks were identified in
both versions of the power spectrum and, not surprisingly, many were only
above SNR $= 2.5$ in the noise-optimized version.  However, four peaks that
were only significant in the sidelobe-optimized version lie close to the
oscillation ridges (three with $l=0$ and one with $l=1$).  The detection of
these peaks vindicates our decision to examine the sidelobe-optimized power
spectrum.

\begin{figure}
\epsscale{0.5}
\plotone{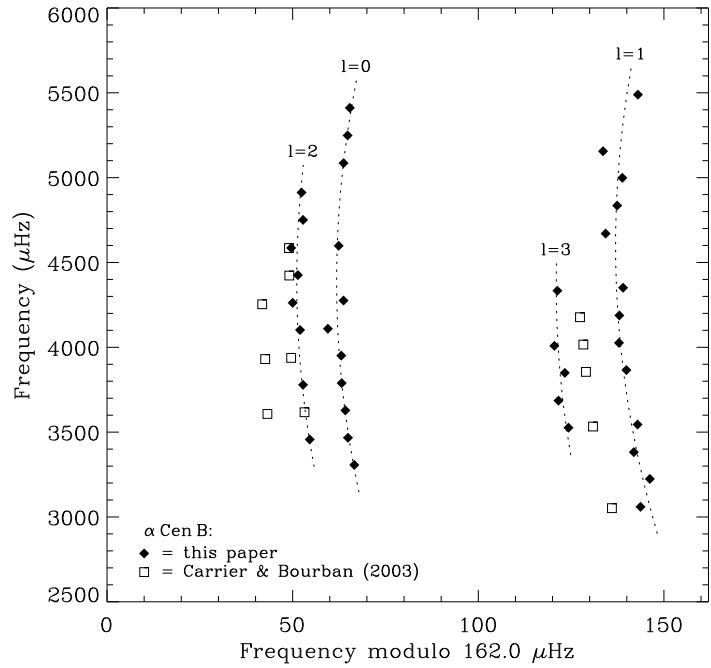}
\caption[]{\label{fig.echelle.final} Echelle diagram of the final
  oscillation frequencies in \acenb{} (filled diamonds), as listed in
  Table~\ref{tab.freqs}.  Dotted curves show the fits to the frequencies,
  given by equations~(\ref{eq.nu0})--(\ref{eq.nu3}).  Open squares indicate
  the frequencies reported by \citet{C+B2003}.}
\end{figure}

The next step was to select those peaks that we believe are due to
oscillations and reject those due to noise.  This is the only subjective
part of the process, but is required if we are to measure as many
frequencies as possible from our data \new{because at this SNR, not all the
extracted peaks will be genuine}.  Our final set of identified modes is
shown in Fig.~\ref{fig.echelle.final} and Table~\ref{tab.freqs} (and also
indicated by the dotted lines in Fig.~\ref{fig.power}).  In some cases the
iterative sine-wave fitting produced two peaks (in one case, three) that
appear to arise from a single oscillation mode, which is to be expected if
we have partially resolved the modes (see Sec.~\ref{sec.lifetimes} for
discussion of the mode lifetimes).  In these cases, and in cases where the
same peak was detected in both versions of the power spectrum, all peaks
are shown in Fig.~\ref{fig.echelle} but they are combined into a single
unweighted mean in Fig.~\ref{fig.echelle.final} and Table~\ref{tab.freqs}
(the uncertainties given in the table are discussed below in
Sec.~\ref{sec.lifetimes}).  We have been conservative in not selecting
three peaks outside the main region because we cannot be sure of the
curvature of the extrapolated ridge lines.  These are: 2827.7\,\muHz{}
($l=0$?), 5894.9\,\muHz{} ($l=2$?) and 5973.1\,\muHz{} ($l=1$?).

\new{Inspection of Fig.~\ref{fig.power} shows some apparent mismatches
between the peaks in the power spectrum and the dotted lines representing
the extracted frequencies.  For example, there are enhancements in power at
3600, 3700 and 4050\,\muHz{} that do not correspond to identified modes.
We noted similar structures in our analysis of \acena{} \citep{BKB2004} and
the explanation is the same: the interaction of the window function with
noise peaks and oscillation peaks.  We have verified that the same
phenomenon occurs in solar data by analysing segments of the
publically-available\footnote{\tt http://www.medoc-ias.u-psud.fr/golf/}
805-day series of full-disk velocity observations taken by the GOLF
instrument (Global Oscillations at Low Frequencies) on the SoHO spacecraft
\citep{UGR2000}, which have a sampling time of 80\,s.  We applied the same
sampling window as our observations of \acenb{} and the results showed
low-level structure in the power spectrum similar to those in
Fig.~\ref{fig.power}.  This confirms that such a structure is a natural
consequence of the spectral window interacting with multi-mode oscillations
having finite lifetimes.}

In Fig.~\ref{fig.echelle.final} the open squares show the frequencies
reported by \citet{C+B2003}.  It is clear that, while they found the
correct large separation, there is a shift in their frequencies of one
cycle per day (11.57\,\muHz).  This reflects the problem associated with
single-site observations, especially when the peaks have low
signal-to-noise.

As with \acena{}, we note a scatter of the peaks about the ridge lines that
is much higher than expected from SNR considerations and which we interpret
as being due to the finite lifetime of the modes.  We therefore fit to the
ridges, in order to obtain more accurate estimates for the eigenfrequencies
of the star.  In the case of \acena{} \citep{BKB2004} we fitted to the
frequencies in a two-step process by first fitting the three small
separations and then fitting a parabola to the individual frequencies.
Here, we adopted a simpler approach that gives almost identical results, in
which we fitted directly to the frequencies \newnew{(see \citealt{Ulr86})}.
The nine fitted parameters specify: the curvatures of the parabolas (one
common value), the large separation at some reference frequency for
each~$l$ (four values) and the absolute positions of the ridges (four
values).  The equations for this fit are as follows:
\begin{eqnarray}
 \nu_{n,0} & = & \left[3950.57 + 161.45 \;(n - 23) + 0.101 \;(n-23)^2\right]\,\muHz \label{eq.nu0}\\
 \nu_{n,1} & = & \left[4026.23 + 161.28 \;(n - 23) + 0.101 \;(n-23)^2\right]\,\muHz \label{eq.nu1}\\
 \nu_{n,2} & = & \left[4101.41 + 161.63 \;(n - 23) + 0.101 \;(n-23)^2\right]\,\muHz \label{eq.nu2}\\
 \nu_{n,3} & = & \left[4171.13 + 161.76 \;(n - 23) + 0.101 \;(n-23)^2\right]\,\muHz.\label{eq.nu3} 
\end{eqnarray}
We show this fit as the dotted curves in Figs.~\ref{fig.echelle}
and~\ref{fig.echelle.final}.

\begin{figure}
\epsscale{0.5}
\plotone{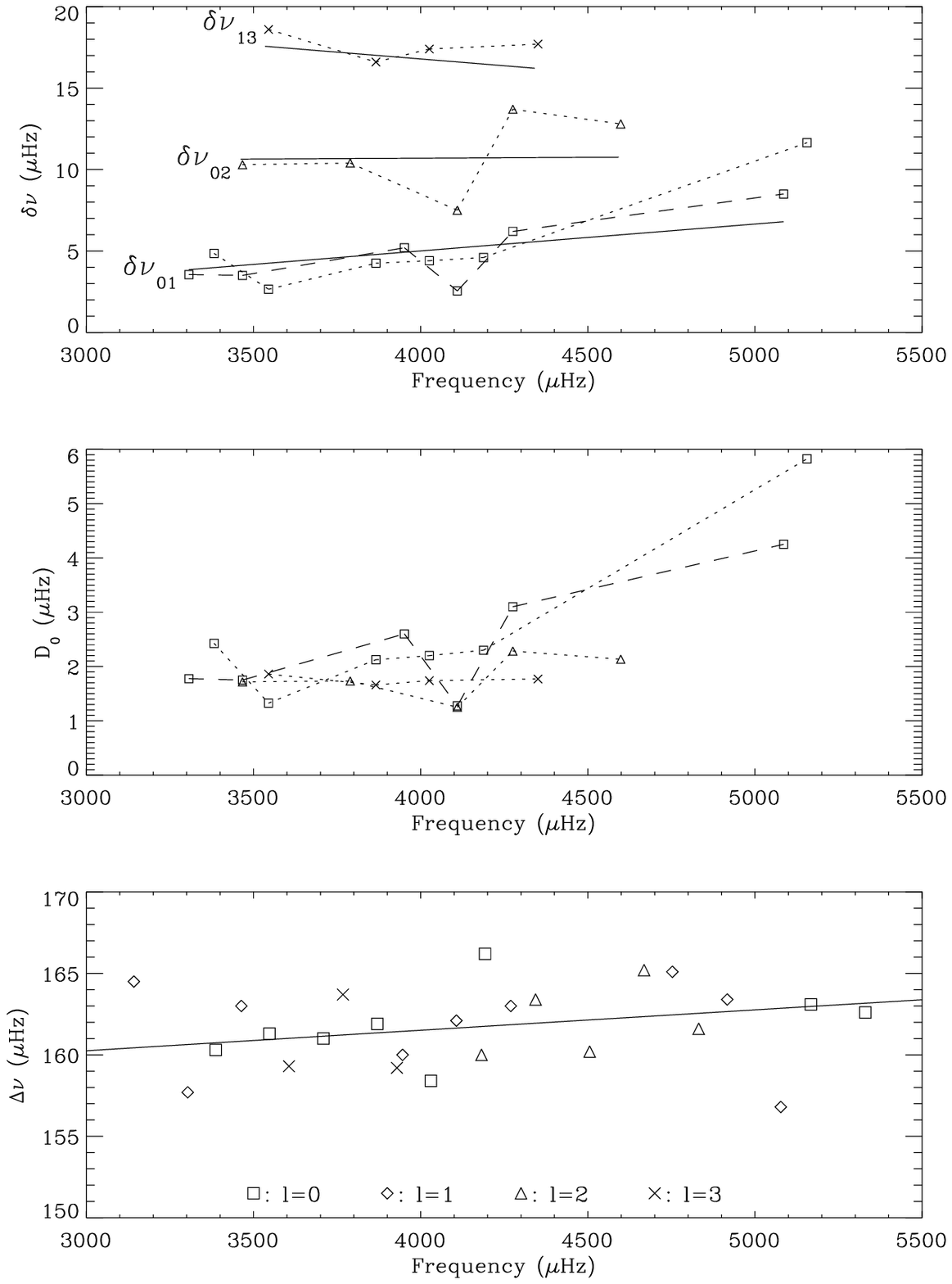}
\caption[]{\label{fig.separations} Frequency separations for \acenb.  Top
  panel: small frequency separations, including $\delta\nu_{01}$ from both
  equations~(\ref{eq.dnu01a}) and~(\ref{eq.dnu01b}) (the dashed line connects
  the latter values).  Solid lines show the separations calculated from
  equations~(\ref{eq.nu0})--(\ref{eq.nu3}).  Middle panel: the $D_0$
  parameter, calculated as $\sixth\dnu{02}$, $\tenth\dnu{13}$ and
  $\half\dnu{01}$, with symbols (and line styles) showing the corresponding
  separations in the top panel.  Bottom panel: the large separations for
  each value of $l$, with the solid line showing $\Delta\nu_0$ as
  calculated from equation~(\ref{eq.nu0}).  }
\end{figure}

Figure~\ref{fig.separations} shows the small frequency separations (top
panel), the $D_0$ parameter (middle panel) and the large separations
($\Delta\nu$; bottom panel), using the same definitions as \citet{BKB2004}.
Thus, $\dnu{02}$ is the separation between adjacent peaks with $l=0$ and
$l=2$ and $\dnu{13}$ is the separation between $l=1$ and $l=3$.  The third
small separation, $\dnu{01}$, is the amount by which $l=1$ modes are offset
from the midpoint between the $l=0$ modes on either side:
\begin{equation}
  \dnu{01} = \half\left(\nu_{n,0} + \nu_{n+1,0}\right) - \nu_{n,1}.
        \label{eq.dnu01a}
\end{equation}
Since one could equally well define $\dnu{01}$ to be the offset of $l=0$
modes from the midpoint between consecutive $l=1$ modes, 
\begin{equation}
  \dnu{01} = \nu_{n,0} - \half\left(\nu_{n-1,1} + \nu_{n,1}\right),
        \label{eq.dnu01b}
\end{equation}
we have shown both versions in Fig.~\ref{fig.separations} (upper panel).
Note that $D_0$, which is sensitive to the stellar core, corresponds to
$\sixth\dnu{02}$, $\tenth\dnu{13}$ and $\half\dnu{01}$ and is a constant if
the asymptotic relation holds exactly.  The solid lines in the top and
bottom panels show the separations calculated from the fitted equations (in
the bottom panel, only $\Delta\nu_0$ is shown; lines for other $l$ values
are almost indistinguishable).  The upward trend of $\Delta\nu$ with
frequency is responsible for the curvature in the echelle diagram.

In Table~\ref{tab.splittings} we give the large and small frequency
separations for \acenb{} at a frequency of 4.0\,mHz.  We also give
$\epsilon$, which is a dimensionless quantity commonly used to parameterize
the absolute position of the frequency spectrum -- see
equation~(\ref{eq.asymptotic}).  We are not able to detect any
statistically significant variations in the large separation with angular
degree ($l$); the weighted mean value ($\Delta\nu$) has an uncertainty of
0.04\%.  We also find that the value of $D_0$ (as defined above) at
4.0\,mHz is the same within uncertainties for all three small separations,
and our mean value for $D_0$ has an uncertainty of 3\%.  We can therefore
place \acenb{} in the so-called asteroseismic H-R diagram (\citealt{ChD84};
see also \new{\citealt{Ulr86};} \citealt{Gou2003,OFChDT2005}), and these
frequency separations should be compared with theoretical models.

Detailed modelling of the \acen{} system was carried out by
\citet{ECT2004}.  They determined the parameters of the system through a
least-squares fit to the observed quantities, including both `classical'
photometric and spectroscopic data and oscillation frequencies from
\citet{B+C2002} and \citet{C+B2003}.  Their two preferred models (Models M1
and M2) both have an age of 6.5\,Gyr.  For \acenb{} they obtained large
separations $\Delta \nu_0$ of 161.7 and 161.1\,\muHz, respectively, while
the $l = 0$ to 2 small separations were $\delta \nu_{02} = 10.3$ and
$10.2\,\muHz$, respectively; both are essentially consistent with the
values found here (cf.\ Table~\ref{tab.splittings}), although with some
slight preference for Model M1.  It should be noted, however, that the
average values of these quantities depend on the detailed way in which the
averages are computed, including the selection of modes.  Also, the
computed values of $\Delta \nu_0$, and to a lesser extent $\delta
\nu_{02}$, are affected by the uncertain physics of the near-surface layers
of the model.  An analysis of the combined set of frequencies for \acena,
presented by \citet{BKB2004}, and the results of \acenb{} shown in
Table~\ref{tab.freqs} is in progress.

\subsection{Curvature at high and low frequencies}  \label{sec.ridges}

We now describe a method that allows us to measure the oscillations at low
SNR, outside the central region in which we can identify individual modes.
This allows us to examine curvature in the echelle diagram, which
corresponds to measuring the large separation as a function of frequency.
The method relies on smoothing the power spectrum in order to increase the
contrast between the oscillation signal and the background noise.  The
first step was to smooth the power spectrum quite heavily, with a FWHM of
$\Delta\nu/4$.  This type of smoothing was used by \citet{GPTC98} to find
high-frequency peaks in the solar power spectrum, but here we use an even
broader smoothing function.  The smoothing combines the power from each
even-degree pair of modes ($l=0$ and 2) into a single resolution element,
and the same applies to the odd-degree pairs ($l=1$ and~3).

\begin{figure}
\epsscale{0.3}
\plotone{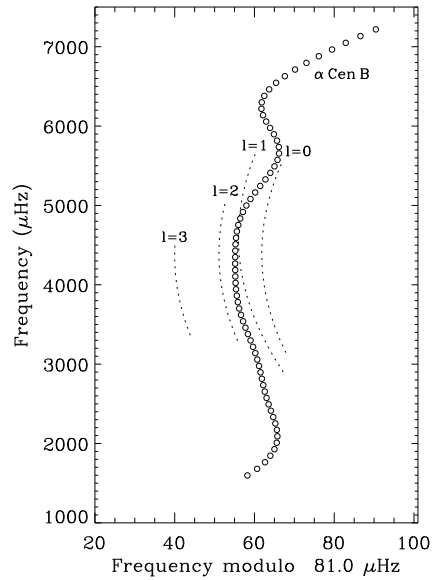}
\caption[]{\label{fig.echelle.smoothed} Smoothed echelle diagram of the
power spectrum of \acenb, plotted with frequencies reduced modulo half the
large separation (see Sec.~\ref{sec.ridges} for details).  The dotted lines
correspond to the fits shown in Fig.~\ref{fig.echelle.final} and given by
equations~(\ref{eq.nu0})--(\ref{eq.nu3}).}
\end{figure}

We next arranged the smoothed power spectrum in echelle format, with
frequencies reduced modulo $\half\Delta\nu$ rather than the conventional
$\Delta\nu$.  This causes power associated with odd and even pairs to line
up, allowing us to smooth in the vertical direction in this echelle diagram
to improve the contrast further (for this smoothing we used FWHM =
$3.4\Delta\nu$).  Finally, we measured the highest peak in each half-order
and these are marked by the open circles in
Fig.~\ref{fig.echelle.smoothed}.  We see a clear ridge of power that
extends well beyond the central region.  The dotted lines are the fits
shown in Fig.~\ref{fig.echelle.final} and given by
equations~(\ref{eq.nu0})--(\ref{eq.nu3}).  We see good agreement in the
central region, but the smoothed data allow us to measure curvature well
outside the region in which we were able to identified individual
oscillation modes.

\begin{figure}
\epsscale{0.5}
\plotone{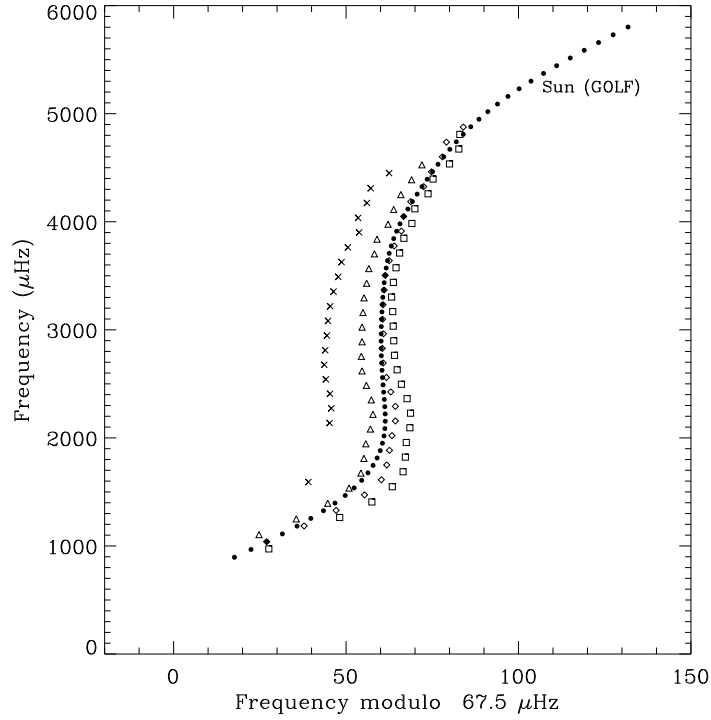}
\caption[]{\label{fig.echelle.solar} Similar to
Fig.~\ref{fig.echelle.smoothed}, but for a 805-day series of GOLF
observations of the Sun.  The open symbols are published solar frequencies
for different $l$ values (open symbols have same meaning as in the bottom
panel of Fig.~\ref{fig.separations}).  }
\end{figure}

In order to evaluate this method, we have also applied it to the
oscillations in the Sun by analysing the 805-day GOLF series (see
Sec.~\ref{sec.freqs}).  We used the same amount of smoothing as for
\acenb{} (when specified in terms of $\Delta\nu$) and the filled circles in
Fig.~\ref{fig.echelle.solar} show the highest peak in each half-order.  The
figure also shows published solar frequencies for $l=0$--3
\citep{LBB97,BVU2000,CEI2002b}.  We see that the ridge of power follows the
published frequencies very well, but again extends beyond them at high
frequencies.  It is interesting that our measurements at high frequency
match up perfectly with the so-called pseudo-modes, also called HIPs
(High-frequency Interference Peaks), which have been seen in smoothed solar
power spectra up to 7.5\,mHz \citep{GPTC98,CEI2001b,GLG2002}.  This may be
relevant to the discussion of the physical nature of these HIPs: are they
related to ordinary resonant p modes, despite having frequencies above the
acoustic cut-off frequency in the atmosphere, and caused by small but
non-zero reflectivity at the photosphere \citep{B+Gough90}; or are they
resonances between direct and reflected waves from a localized source
\citep[e.g.,][]{KDH90,K+L91}?

\begin{figure}
\epsscale{0.5}
\plotone{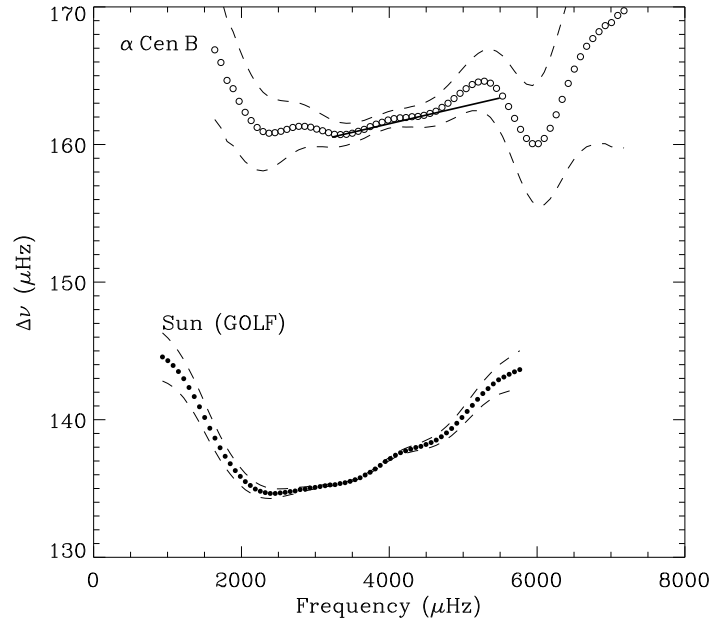}
\caption[]{\label{fig.large.sep} The large separation as a function of
  frequency in \acenb{} and the Sun, as derived from
  Figs.~\ref{fig.echelle.smoothed} and~\ref{fig.echelle.solar}.  The dashed
  lines indicate the $\pm1\sigma$ uncertainties on $\Delta\nu$ (see
  text). The solid line for \acenb{} is the same as in the bottom panel of
  Fig.~\ref{fig.separations} and shows $\Delta\nu_0$ as calculated from
  equation~(\ref{eq.nu0}).}
\end{figure}

The points in Fig.~\ref{fig.large.sep} show the large separation as a
function of frequency for both \acenb{} and the Sun.  These values were
derived from Figs.~\ref{fig.echelle.smoothed} and~\ref{fig.echelle.solar},
respectively, simply by doubling the differences between consecutive points
along the ridges.  The two curves are remarkably similar, except for the
pronounced dip at 6\,mHz in \acenb{}, which occurs in the region of low S/N
and which we suspect is not real.  In fact, our tests show that this type
of feature sometimes appears as an artifact when the method is applied to
short segments of the solar GOLF data (to match the observing window of
\acenb{}).  The same tests also show that the method does measure
$\Delta\nu$ very accurately in the regions where oscillations have
significant amplitude.  To illustrate this, the dashed lines in
Fig.~\ref{fig.large.sep} show our best estimates, based on these tests, of
the $\pm1\sigma$ uncertainties in $\Delta\nu$.  We conclude that this is a
powerful technique for measuring the large separation over an extended
frequency range.  For \acenb{} this will provide extra constraints on
theoretical models while, for the Sun, we have been able to measure
$\Delta\nu$ to much higher frequencies than has been done previously.
\newnew{The S-shaped structure in the echelle diagram indicates a departure
from the second-order fit used in equations~(\ref{eq.nu0})--(\ref{eq.nu3}),
implying that cubic (or higher) terms are needed to describe the
frequencies fully.  As pointed out by \citet{Ulr88}, these higher-order
coefficients might be useful in constraining the stellar mass.}

\section{Mode lifetimes}        \label{sec.lifetimes}

The scatter of the observed frequencies about the ridges allows us to
estimate the mode lifetimes.  In the case of \acena{} we did this by
measuring the deviations of the measured frequencies from the fitted
relations and comparing with simulations \citep{BKB2004}.  Here we adopted
a slightly different approach that has the advantage of being independent of
the fit, although it does still assume that we have correctly assigned $n$
and~$l$ values.  For each measured frequency~$\nu_{n,l}$ (solid points in
Fig.~\ref{fig.echelle.final}), except those at the ends of the ridges, we
calculated the difference~$\Delta_{n,l}$ between the measured frequency and
that expected from the positions of the two nearest neighbours on the same
ridge, using linear interpolation:
\begin{equation}
  \Delta_{n,l} =  \nu_{n,l} \; - \; \frac{\Delta n_- \; \nu_{(n+\Delta n_+),l} 
    \;+\; \Delta n_+
    \; \nu_{(n-\Delta n_-),l}}{\Delta n_+ + \Delta n_-}.
\end{equation}
Here, the two neighbouring modes lie in orders $n+\Delta n_+$ and $n-\Delta
n_-$, respectively.  In most cases, $\Delta n_-$ and $\Delta n_+$ are~1,
but in some cases one or both is 2 or even~3.  We therefore need to convert
$\Delta_{n,l}$ to a quantity that we can easily compare with simulations.
We choose this to be the rms scatter of that peak about its expected
position, calculated as follows:
\begin{equation}
  \sigma^2_{n,l} =   \frac{(\Delta_{n,l})^2 \; (\Delta n_- + \Delta n_+)^2}{(\Delta
  n_- + \Delta n_+)^2 + (\Delta n_-)^2 + (\Delta n_+)^2}.
\end{equation}

\begin{figure}
\epsscale{0.5}
\plotone{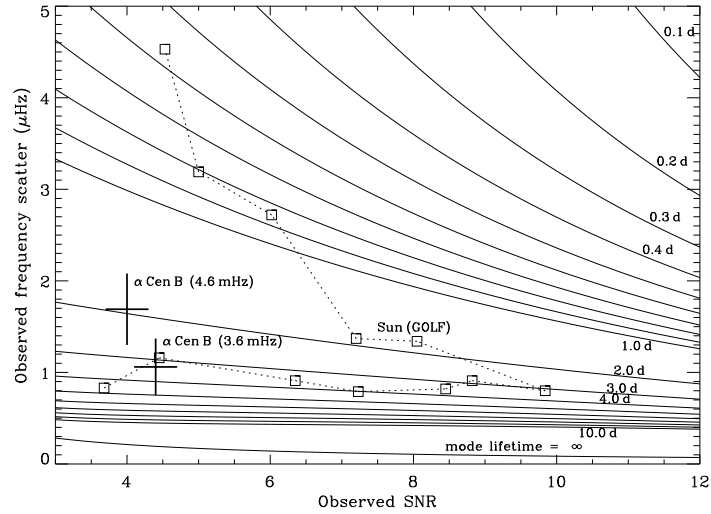}
\caption[]{\label{fig.lifetime.snr} Calibration of mode lifetimes for the
  \acenb{} observing window, using the noise-optimized weights.  Solid
  lines are the results of simulations and show frequency scatter versus
  SNR for various mode lifetimes.  Crosses show actual results for \acenb{}
  in two frequency ranges, while open squares show results from GOLF data
  for twelve consecutive $l=1$ modes in the Sun.}
\end{figure}

Averaging $\sigma$ over many peaks gives an estimate of the mode lifetime,
but we must keep in mind that the finite SNR also introduces a scatter to
the peak positions.  To calibrate these two contributions, we carried out a
large number of simulations (175\,000), each with a single input frequency
and each sampled with our observational window function (with the
noise-optimized weights).  We used the method described by \citet{SKB2004}
to generate the time series of an oscillation that was re-excited
continuously with random kicks and damped on a timescale that was an
adjustable parameter (the mode lifetime).  The other adjustable input
parameter was the oscillation amplitude, while the mode frequency and the
noise level were fixed.  We made 100 simulations for each set of input
parameters and from the resulting power spectra we measured the rms scatter
in frequency of the highest peak and its mean SNR (discarding peaks with
frequencies more than $4\sigma$ from actual value).  We repeated this for
various values of the input parameters and the results are summarised in
Fig.~\ref{fig.lifetime.snr}.  Each solid line in this figure shows the
observed frequency scatter versus the observed SNR for a given value of the
mode lifetime.  As expected, the frequency scatter increases both with
decreasing mode lifetime and with decreasing SNR\@.  We stress that this
figure applies specifically to our observing window for \acenb{} and should
be recalculated for other observing windows.

The crosses in Fig.~\ref{fig.lifetime.snr} show our results for \acenb{} in
two frequency ranges, centred at 3.6\,mHz (the mean of all peaks below
4.0\,mHz) and at 4.6\,mHz (the mean of peaks above 4.0\,mHz).  From these
crosses we can determine the mode lifetimes.  In order to check our
results, we have also analysed segments of the same 805-day GOLF series of
full-disk velocity observations of the Sun used in Sec.~\ref{sec.ridges}.
We divided this series into 100 segments, and imposed on each the \acenb{}
window function (with weights).  The squares in Fig.~\ref{fig.lifetime.snr}
show the frequency scatter for each $l=1$ mode as a function of the mean
SNR (both measured over the 100 segments).  We analysed twelve $l=1$ modes
($n=16$--27) and ignored the pairs with $l=0$ and~$2$ because they are less
well resolved and interact via daily sidelobes.  We see the well-known
result that mode lifetimes in the Sun vary with frequency (the shortest
lifetimes occur at the highest frequencies).  Those with intermediate
frequency have the best SNR because they have the highest amplitudes.

For each measured point in Fig.~\ref{fig.lifetime.snr} (two for \acenb{}
and twelve for the Sun), we used our simulations to infer the mode
lifetime.  The results are shown as a function of frequency in
Fig.~\ref{fig.lifetime.freq} using the same symbols, where we have
expressed frequency in units of~$\nu_{\rm max}$, the frequency of maximum
oscillation power (see Table~\ref{tab.amps} for values of~$\nu_{\rm max}$).
The diamonds in this figure are published measurements of the solar mode
lifetimes \citep{CEI97}, and we see good agreement with our values.  We
also see that the typical mode lifetimes for \acenb{}, when considered as a
function of $\nu/\nu_{\rm max}$, are similar to those in the Sun.  Our
estimates are 3.3$^{+1.8}_{-0.9}$\,d at 3.6\,mHz and 1.9$^{+0.7}_{-0.4}$\,d
at 4.6\,mHz.

\subsection{Revised mode lifetimes for \acena}

\begin{figure}
\epsscale{0.5}
\plotone{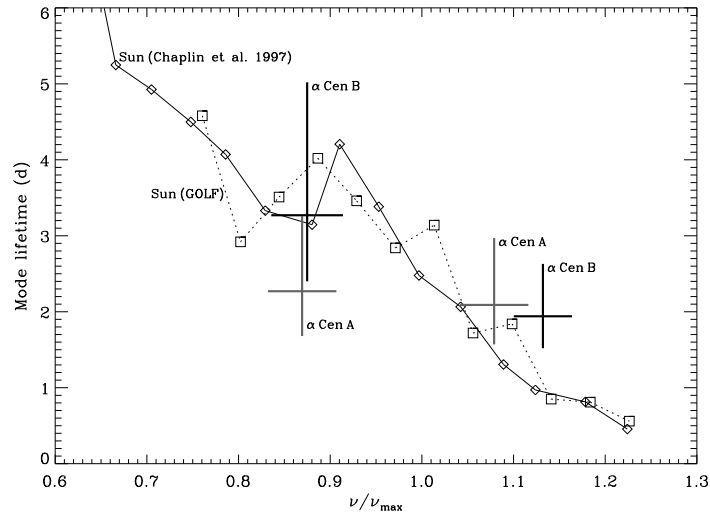}
\caption[]{\label{fig.lifetime.freq} Mode lifetimes versus normalized
  frequency, where $\nu_{\rm max}$ is the frequency of maximum oscillation
  power.  Values for \acenb{} (black crosses) and the Sun observed by GOLF
  (open squares) are calculated from Fig.~\ref{fig.lifetime.snr}.  Two
  values for \acena{} (grey crosses) are calculated from a similar
  calibration of the results presented by \citet{BKB2004}. Open diamonds
  are published measurements of the solar mode lifetimes \citep{CEI97}.  }
\end{figure}

The grey crosses in Fig.~\ref{fig.lifetime.freq} show mode lifetimes for
\acena{}, based on the observations analysed by \citet{BKB2004}.  Here, we
have re-measured the mode lifetimes using the method described above, which
involved making a whole new series of simulations with the \acena{} window
function in order to convert frequency scatters into lifetimes.  The
inferred lifetimes are 2.3$^{+1.0}_{-0.6}$\,d at 2.1\,mHz and
2.1$^{+0.9}_{-0.5}$\,d at 2.6\,mHz.

We can compare these revised lifetimes for \acena{} with those we reported
previously from the same data \citep[see Table~3 in][]{BKB2004}:
1.4$^{+0.5}_{-0.4}$\,d at 2.1\,mHz and 1.3$^{+0.5}_{-0.4}$\,d at 2.6\,mHz.
The revised values are higher, although the $1\sigma$ error bars do
overlap.  The reason for the change is that our previous calculation
underestimated the contribution of SNR to the frequency scatter.  We did of
course include the effects of SNR, but did so by treating it as being
independent of the scatter introduced by finite mode lifetimes.  In fact,
as our new simulations show, the two contributions are not independent.  If
they were, the curves in Fig.~\ref{fig.lifetime.snr} would not rise towards
low SNR as steeply as they do.

The important conclusion, as we can see from Fig.~\ref{fig.lifetime.freq},
is that mode lifetimes in \acena{} are not substantially lower than those
in the Sun, although the value for the lower frequency range is still
about~$1.5\sigma$ below solar.

\section{Amplitudes and noise levels}  \label{sec.amp}

It is important to measure oscillation amplitudes in solar-like stars and
to compare these with theoretical calculations \citep[e.g.,][]{HBChD99}.
It is also interesting to measure the background noise from stellar
convection, although in velocity this requires extremely precise
measurements because the signature is weak.  For both these measurements,
we have chosen to smooth the power spectrum heavily, so as to produce a
single hump of excess power that is insensitive to the fact that the
oscillation spectrum has discrete peaks.  It is also useful to convert to
power density, which is independent of the observing window and therefore
allows us to compare noise levels.  This is done by multiplying the power
by the effective length of the observing run, which we calculated as the
reciprocal of the area under the spectral window (in power).

\begin{figure}
\epsscale{0.5}
\plotone{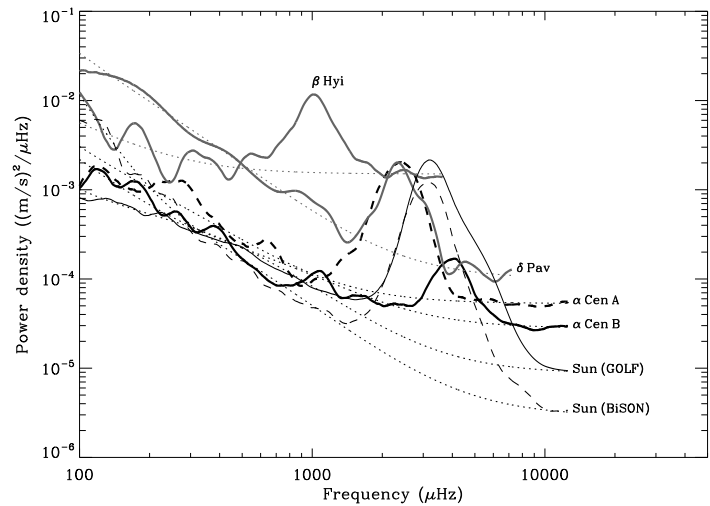}
\caption[]{\label{fig.pd} Smoothed power density spectra from velocity
  observations of the Sun and four other stars.  The dotted lines are fits
  to the noise background.
}
\end{figure}

In Fig.~\ref{fig.pd} we show smoothed power density spectra for the Sun and
four other stars: \acena{} and~B, \dpav{} and \bhyi.  For the four stars,
we used the most precise observations available: UVES observations for
\acenb{} (this paper), \dpav{} (this paper) and \acena{} \citep{BBK2004},
and UCLES observations for \bhyi{} \citep{BBK2001}.  In all cases we used
the raw velocity measurements, before removal of any jumps or slow trends,
since we are interested in measuring the total noise level.

For the Sun, we used data from BiSON (Birmingham Solar Oscillations
Network) and GOLF\@.  The BiSON data comprised a 7-day time series with
40-second sampling from the Las Campanas station in Chile, kindly provided
by W. Chaplin (private comm.).  The GOLF data comprised a 20-day time
series with 20-second sampling, kindly provided by P.~Boumier (private
comm.).  Note that these have a higher Nyquist frequency than the publicly
available GOLF data, which we used in Secs.~\ref{sec.ridges}
and~\ref{sec.lifetimes} but which are only sampled at 80\,s.

The dotted lines in Fig.~\ref{fig.pd} are fits to the noise backgrounds,
based on the \citet{Har84} model of solar granulation.  The Harvey model
gives a convenient functional form, even in stars where the low-frequency
noise has a strong additional contribution from instrumental noise.  We
discuss the noise levels in more detail below.  First, we show that these
smoothed power density spectra provide a powerful way to measure
oscillation amplitudes in a way that is independent of mode lifetime.

To do this, we first subtracted the background noise (dotted lines in
Fig.~\ref{fig.pd}) from each observed power density spectrum.  We only
included those parts of the spectrum that were at least twice the noise
level.  In order to calculate the amplitude per oscillation mode, we should
then multiply by $\Delta\nu/4$ (where $\Delta\nu$ is the large frequency
separation of the star) and take the square root.  The rationale for this
is that there are four modes in each segment of length $\Delta\nu$ (with
$l=0$, 1, 2 and~3).  However, we must keep in mind that modes with
different angular degrees have different visibilities in full-disk
observations, due to varying amounts of cancellation.  Based on the results
presented by \citet{BKR96}, we calculated the effective number of modes per
$\Delta\nu$, normalised to the mean of the $l=0$ and $l=1$ amplitudes, to
be 3.0.  We therefore used this factor, rather than 4, in our calculation.

For $\Delta\nu$ we used the following values: 135\,\muHz{} for the Sun;
106\,\muHz{} for \acena; 162\,\muHz{} for \acenb; 56\,\muHz{} for \bhyi{}
and 93\,\muHz{} for \dpav.  The last of these is not a measurement, since
none is available, and is instead derived from the following adopted
parameters: $M=0.9\,M_\odot$, $L=1.3\,L_\odot$ and $T_{\rm eff} = 5540$\,K.

\begin{figure}
\epsscale{0.5}
\plotone{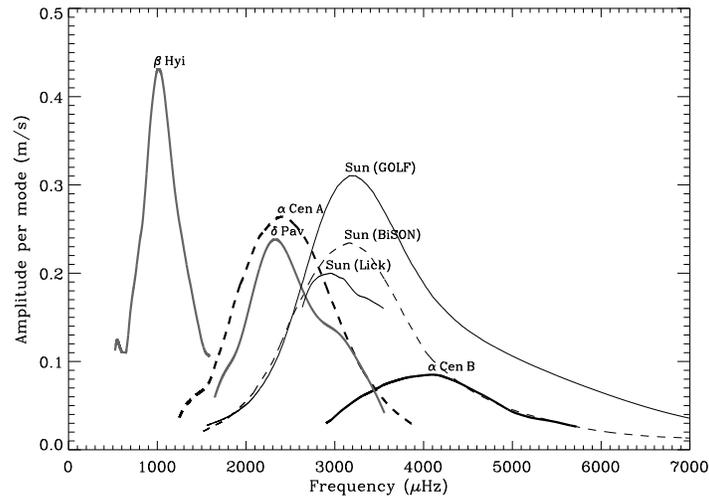}
\caption[]{\label{fig.amps} Amplitude per mode for solar-like oscillations.
  These curves were calculated from those in Fig.~\ref{fig.pd} by
  subtracting the background noise, multiplying by $\Delta\nu/3.0$ and
  taking the square root (see text).  We also show oscillation amplitudes
  in the Sun, measured from iodine-referenced observations at Lick
  Observatory.  }
\end{figure}

Our amplitude estimates are shown in Fig.~\ref{fig.amps}, and the height,
frequency and FWHM of the envelopes are given in Table~\ref{tab.amps}.  We
see a number of interesting things from Figs.~\ref{fig.pd}
and~\ref{fig.amps}.  Looking first at the Sun, the results illustrate very
nicely that the solar oscillation amplitude depends on the spectral line
that is being measured \new{(see \citet{BSG2005} for a recent study of this
phenomenon)}.  The sodium line used by GOLF is formed higher in the solar
atmosphere than the potassium line used by BiSON, which is why GOLF
measures higher oscillation amplitudes \citep{IMP89}.  The actual height
difference is difficult to estimate; \citet{PRRC92} quoted $\sim$200\,km
while \citet{BSG2005} adopted $\sim$60\,km.  Velocity measurements of other
stars are made using a wide wavelength range so as to include many spectral
lines, and the Doppler signal is dominated by neutral iron lines.  Since
these lines are formed about 400\,km below the sodium D lines
\citep{EMR2001,M+K2003}, we would expect solar amplitudes measured using
the iodine technique to be less than those from both BiSON and GOLF\@.

Unfortunately, there do not appear to be any published estimates of the
solar oscillation amplitude using the stellar technique.  Here we present
some previously unpublished observations of the solar spectrum made using
iodine referencing.  One of us (J.T.W.) made observations of the Moon using
the 0.6 m Coude Auxilliary Telescope (CAT) at Lick Observatory, which fed
the Hamilton Spectrometer, a high-resolution ($R=60000$) echelle
\citep{Vog87}.  The CAT tracked a fixed point of uniform surface brightness
(Archimedes crater) for five consecutive nights near full moon, thus
measuring the disk-integrated solar spectrum at night (a technique inspired
by \citealt{McMMP93}).  These velocity measurements allowed us to estimate
the solar oscillation amplitude, which we include in Fig.~\ref{fig.amps}
and Table~\ref{tab.amps}.  The results support the conclusion that Fe~I
measurements give lower amplitudes than both GOLF and BiSON\@.  They also
place the solar amplitude between those of \acena{} and~B, as would be
expected given their stellar parameters.  It would clearly be valuable to
obtain more measurements of the Sun, in order to better calibrate the
relationship between stellar and solar amplitudes.  The method described
here allows us to estimate amplitudes independently of mode lifetime and
observing window.  Comparing the amplitudes of different stars with
theoretical models is the subject of a future paper.  

We turn now to the noise levels in the various observations, looking first
at low frequencies.  The rise in power towards low frequencies seen in
Fig.~\ref{fig.pd} is due to a combination of instrumental drift and stellar
background noise.  Of course, it is very difficult to distinguish between
these two, although in the Sun it is established the solar background is
dominant.  It therefore seems likely that the low-frequency power from
\acena{} is also mostly stellar, given the power density is similar to the
Sun \citep[see also][]{KBF99}.  The same may also be true for \acenb.  We
can certainly say that at 1\,mHz, the granulation noise in both stars is no
greater than is observed in the Sun by GOLF and BiSON\@.

At the highest frequencies the noise levels are dominated by white noise
from photon statistics.  We can see that the power density (which indicates
noise per unit observing time) is lowest for BiSON and GOLF, followed by
\acenb{} (UVES) and \acena{} (UVES).  In the last two columns of
Table~\ref{tab.amps} we provide an update to Table~1 of \citet{BBK2004},
showing the noise \new{per minute of observing} at frequencies just above
the p-mode envelope ($2\nu_{\rm max}$) and also, where the sampling allows,
at very high frequencies (11\,mHz).  We should note that the power at high
frequencies in the Sun (4--6\,mHz), particularly in the sodium line used by
GOLF, is dominated by solar noise that presumably arises from chromospheric
effects, with only a small fraction being due to coherent p-modes.

\section{Conclusions}

Our observations of \acenb{} from two sites have allowed us to identify 37
oscillation modes with $l=0$--$3$.  Fitting to these modes gave the large
and small frequency separations as a function of frequency.  We also
introduced a new method, involving smoothing in the $\half\Dnu{}$ echelle
diagram, that allowed us to trace the ridges of power, and hence measure
the large separation, well beyond the central region.

We inferred the mode lifetimes in two frequency ranges by measuring the
scatter of the oscillation frequencies about a smooth trend, based on a
calibration involving extensive simulations.  We found mode lifetimes in
\acenb, when considered as a function of frequency relative to the maximum
power, that are consistent with those seen in the Sun.  We applied the same
analysis to our observations of \acena{} and deduced revised mode lifetimes
that are slightly higher than we previously published \citep{BKB2004}.

A limited set of observations of the star \dpav{} showed oscillations
centred at 2.3\,mHz with peak velocity amplitudes close to solar.  Further
observations are needed to determine the large separation and individual
mode frequencies in this star.

Finally, we also introduced a new method of measuring oscillation
amplitudes from heavily-smoothed power density spectra.  We estimated the
amplitude per mode of \acena~and~B, \bhyi, \dpav{} and the Sun, and pointed
out that the results may depend on which spectral lines are used for the
velocity measurements.

\acknowledgments

We thank Alan Gabriel and Patrick Boumier from the GOLF team for useful
comments and for providing the GOLF data at 20-second sampling.  We also
thank Bill Chaplin for providing data from BiSON and for useful
discussions.  This work was supported financially by the Australian
Research Council, by the Danish Natural Science Research Council and by the
Danish National Research Foundation through its establishment of the
Theoretical Astrophysics Center.  We further acknowledge support by NSF
grant AST-9988087 (RPB), and by SUN Microsystems.

\clearpage

\begin{table*}
\small
\caption{\label{tab.freqs} Oscillation frequencies for \acenb{} (\muHz)}
\begin{center}
\begin{tabular}{ccccc}
\tableline
\tableline
\noalign{\smallskip}
~~~$n$~~~ &  ~~~$l=0$~~~  & ~~~$l=1$~~~  & ~~~$l=2$~~~  & ~~~$l=3$~~~ \\
\tableline
\noalign{\smallskip}
17  &     \ldots        &  $3059.7\pm0.9$ &   \ldots        &   \ldots       \\
18  &     \ldots        &  $3224.2\pm0.9$ &   \ldots        &   \ldots       \\
19  &    $3306.6\pm0.9$ &  $3381.9\pm1.1$ &  $3456.6\pm1.1$ &  $3526.3\pm1.1$\\
20  &    $3466.9\pm1.0$ &  $3544.9\pm0.8$ &   \ldots        &  $3685.6\pm1.1$\\
21  &    $3628.2\pm1.0$ &   \ldots        &  $3778.8\pm1.1$ &  $3849.3\pm1.3$\\
22  &    $3789.2\pm1.2$ &  $3865.9\pm1.1$ &   \ldots        &  $4008.5\pm1.5$\\
23  &    $3951.1\pm1.2$ &  $4025.9\pm1.2$ &  $4102.0\pm1.2$ &   \ldots       \\
24  &    $4109.5\pm1.1$ &  $4188.0\pm1.1$ &  $4262.0\pm1.5$ &  $4333.3\pm1.4$\\
25  &    $4275.7\pm1.5$ &  $4351.0\pm1.4$ &  $4425.4\pm1.5$ &   \ldots       \\
26  &     \ldots        &   \ldots        &  $4585.6\pm1.8$ &   \ldots       \\
27  &    $4598.4\pm1.6$ &  $4670.3\pm1.7$ &  $4750.8\pm1.8$ &   \ldots       \\
28  &     \ldots        &  $4835.4\pm2.0$ &  $4912.4\pm2.0$ &   \ldots       \\
29  &     \ldots        &  $4998.8\pm1.9$ &   \ldots        &   \ldots       \\
30  &    $5085.7\pm2.2$ &  $5155.6\pm2.1$ &   \ldots        &   \ldots       \\
31  &    $5248.8\pm2.4$ &   \ldots        &   \ldots        &   \ldots       \\
32  &    $5411.4\pm1.9$ &  $5489.0\pm2.3$ &   \ldots        &   \ldots       \\
\noalign{\smallskip}
\tableline
\end{tabular}
\end{center}
\end{table*}

\begin{table*}
\small
\caption{\label{tab.splittings}  Frequency separations at 4.0\,{\rm mHz}
  for \acenb{} (\muHz)} 
\begin{center}
\begin{tabular}{lrr}
\tableline
\tableline
\noalign{\smallskip}
$\Delta\nu_0$  &      $161.50 \pm 0.11$  &  (0.07\%)\\
$\Delta\nu_1$  &      $161.27 \pm 0.09$  &  (0.06\%)\\
$\Delta\nu_2$  &      $161.48 \pm 0.17$  &  (0.11\%)\\
$\Delta\nu_3$  &      $161.53 \pm 0.33$  &  (0.20\%)\\
\noalign{\smallskip}
$\Delta\nu$    &      $161.38 \pm 0.06$  &  (0.04\%)\\
\noalign{\smallskip}
\tableline
\noalign{\smallskip}
$\delta\nu_{01}$ & $ 4.52  \pm  0.51 $ & (11\%)\\
$\delta\nu_{02}$ & $10.14  \pm  0.62 $ &  (6\%)\\
$\delta\nu_{13}$ & $16.73  \pm  0.65 $ &  (4\%)\\    
$D_0$            & $ 1.771 \pm  0.04\rlap{7}$ &  (3\%)\\
\noalign{\smallskip}
\tableline
\noalign{\smallskip}
$\epsilon$       & $ 1.477 \pm  0.01\rlap{1}$ &  (0.7\%)\\
\noalign{\smallskip}
\tableline
\end{tabular}
\end{center}
\end{table*}

\begin{table*}
\begin{center}
\caption{\label{tab.amps}Amplitudes and noise levels for solar-like oscillations}
\small
\begin{tabular}{lcccccc}
\hline
\hline
     &             & Peak amplitude & $\nu_{\rm max}$ & FWHM  &\multicolumn{2}{c}{Noise per min (\ms)} \\
Star & Spectrograph & per mode (\ms) & (mHz)           & (mHz) & at $2\,\nu_{\rm
  max}$  & at 11\,mHz         \\
\hline
\noalign{\smallskip}
\acenb & UVES  & $0.085\pm0.004$ & $4.09 \pm0.17 $ & $1.98\pm0.14 $ & 0.35 & 0.32 \\
\acena & UVES  & $0.263\pm0.008$ & $2.41 \pm0.13 $ & $1.34\pm0.04 $ & 0.49 & 0.45 \\
\bhyi  & UCLES & $0.432\pm0.016$ & $1.02 \pm0.05 $ & $0.54\pm0.05 $ & 2.47 & ---  \\
\dpav  & UVES  & $0.236\pm0.008$ & $2.33 \pm0.09 $ & $1.24\pm0.06 $ & 0.82 & ---  \\
Sun    & GOLF  & $0.308\pm0.005$ & $3.21 \pm0.11 $ & $1.76\pm0.02 $ & 0.52 & 0.20 \\
Sun    & BiSON & $0.233\pm0.006$ & $3.17 \pm0.13 $ & $1.62\pm0.04 $ & 0.20 & 0.12 \\
Sun    & Lick  & $0.208\pm0.029$ & $2.95 \pm0.31 $ & ---            & ---  & ---  \\
\noalign{\smallskip}
\hline
\end{tabular}
\end{center}
\end{table*}


\begin{thebibliography}{38}
\expandafter\ifx\csname natexlab\endcsname\relax\def\natexlab#1{#1}\fi
\expandafter\ifx\csname url\endcsname\relax
  \def\url#1{{\tt #1}}\fi

\bibitem[Balmforth \& {Gough}(1990)Balmforth, \& {Gough}]{B+Gough90}
Balmforth, N.~J., \& {Gough}, D.~O., 1990, ApJ, 362, 256.

\bibitem[Baudin et~al.(2005)Baudin, {Samadi}, {Goupil}, {Appourchaux},
  {Barban}, {Boumier}, {Chaplin}, \& {Gouttebroze}]{BSG2005}
Baudin, F., {Samadi}, R., {Goupil}, M.-J., {Appourchaux}, T., {Barban}, C.,
  {Boumier}, P., {Chaplin}, W.~J., \& {Gouttebroze}, P., 2005, A\&A, 433, 349.

\bibitem[Bedding et~al.(2001)Bedding, Butler, Kjeldsen, Baldry, O'Toole,
  Tinney, Marcy, Kienzle, \& Carrier]{BBK2001}
Bedding, T.~R., Butler, R.~P., Kjeldsen, H., Baldry, I.~K., O'Toole, S.~J.,
  Tinney, C.~G., Marcy, G.~W., Kienzle, F., \& Carrier, F., 2001, ApJ, 549,
  L105.

\bibitem[Bedding et~al.(2004)Bedding, Kjeldsen, Butler, McCarthy, Marcy,
  O'Toole, Tinney, \& Wright]{BKB2004}
Bedding, T.~R., Kjeldsen, H., Butler, R.~P., McCarthy, C., Marcy, G.~W.,
  O'Toole, S.~J., Tinney, C.~G., \& Wright, J.~T., 2004, ApJ, 614, 380.

\bibitem[Bedding et~al.(1996)Bedding, Kjeldsen, Reetz, \& Barbuy]{BKR96}
Bedding, T.~R., Kjeldsen, H., Reetz, J., \& Barbuy, B., 1996, MNRAS, 280, 1155.

\bibitem[Bertello et~al.(2000)Bertello, {Varadi}, {Ulrich}, {Henney},
  {Kosovichev}, \& {Garc{\'{\i}}a}]{BVU2000}
Bertello, L., {Varadi}, F., {Ulrich}, R.~K., {Henney}, C.~J., {Kosovichev},
  A.~G., \& {Garc{\'{\i}}a}, 2000, ApJ, 537, L143.

\bibitem[Bouchy \& {Carrier}(2001)Bouchy, \& {Carrier}]{B+C2001}
Bouchy, F., \& {Carrier}, F., 2001, A\&A, 374, L5.

\bibitem[Bouchy \& {Carrier}(2002)Bouchy, \& {Carrier}]{B+C2002}
Bouchy, F., \& {Carrier}, F., 2002, A\&A, 390, 205.

\bibitem[Butler et~al.(2004)Butler, Bedding, Kjeldsen, McCarthy, O'Toole,
  Tinney, Marcy, \& Wright]{BBK2004}
Butler, R.~P., Bedding, T.~R., Kjeldsen, H., McCarthy, C., O'Toole, S.~J.,
  Tinney, C.~G., Marcy, G.~W., \& Wright, J.~T., 2004, ApJ, 600, L75.

\bibitem[Butler et~al.(1996)Butler, Marcy, Williams, McCarthy, Dosanjh, \&
  Vogt]{BMW96}
Butler, R.~P., Marcy, G.~W., Williams, E., McCarthy, C., Dosanjh, P., \& Vogt,
  S.~S., 1996, PASP, 108, 500.

\bibitem[Carrier \& {Bourban}(2003)Carrier, \& {Bourban}]{C+B2003}
Carrier, F., \& {Bourban}, G., 2003, A\&A, 406, L23.

\bibitem[Chaplin et~al.(2001)Chaplin, {Elsworth}, {Isaak}, {Marchenkov},
  {Miller}, \& {New}]{CEI2001b}
Chaplin, W.~J., {Elsworth}, Y., {Isaak}, G.~R., {Marchenkov}, K.~I., {Miller},
  B.~A., \& {New}, R., 2001, In {\em Helio- and Asteroseismology at the Dawn of
  the Millenium, Proc. SOHO 10/GONG 2000 Workshop}, ESA SP-464, page 191.

\bibitem[Chaplin et~al.(2002)Chaplin, {Elsworth}, {Isaak}, {Marchenkov},
  {Miller}, {New}, {Pinter}, \& {Appourchaux}]{CEI2002b}
Chaplin, W.~J., {Elsworth}, Y., {Isaak}, G.~R., {Marchenkov}, K.~I., {Miller},
  B.~A., {New}, R., {Pinter}, B., \& {Appourchaux}, T., 2002, MNRAS, 336, 979.

\bibitem[Chaplin et~al.(1997)Chaplin, {Elsworth}, {Isaak}, {McLeod}, {Miller},
  \& {New}]{CEI97}
Chaplin, W.~J., {Elsworth}, Y., {Isaak}, G.~R., {McLeod}, C.~P., {Miller},
  B.~A., \& {New}, R., 1997, MNRAS, 288, 623.

\bibitem[Christensen-Dalsgaard(1984)]{ChD84}
Christensen-Dalsgaard, J., 1984, In Mangeney, A., \& Praderie, F., editors,
  {\em Workshop on Space Research in Stellar Activity and Variability},
  page~11. Meudon: Observatoire de Paris.

\bibitem[Deeming(1975)]{Dee75}
Deeming, T.~J., 1975, Ap\&SS, 36, 137.

\bibitem[Eggenberger et~al.(2004)Eggenberger, {Charbonnel}, {Talon}, {Meynet},
  {Maeder}, {Carrier}, \& {Bourban}]{ECT2004}
Eggenberger, P., {Charbonnel}, C., {Talon}, S., {Meynet}, G., {Maeder}, A.,
  {Carrier}, F., \& {Bourban}, G., 2004, A\&A, 417, 235.

\bibitem[Eibe et~al.(2001)Eibe, {Mein}, {Roudier}, \& {Faurobert}]{EMR2001}
Eibe, M.~T., {Mein}, P., {Roudier}, T., \& {Faurobert}, M., 2001, A\&A, 371,
  1128.

\bibitem[Frandsen et~al.(1995)Frandsen, Jones, Kjeldsen, Viskum, Hjorth,
  Andersen, \& Thomsen]{FJK95}
Frandsen, S., Jones, A., Kjeldsen, H., Viskum, M., Hjorth, J., Andersen, N.~H.,
  \& Thomsen, B., 1995, A\&A, 301, 123.

\bibitem[Garc{\'{\i}}a et~al.(1998)Garc{\'{\i}}a, Pall{\'e}, Turck-Chi{\`e}ze,
  {Osaki}, {Shibahashi}, {Jeffries}, {Boumier}, {Gabriel}, {Grec}, {Robillot},
  Roca~Cort{\'e}s, \& {Ulrich}]{GPTC98}
Garc{\'{\i}}a, R.~A., Pall{\'e}, P.~L., Turck-Chi{\`e}ze, S., {Osaki}, Y.,
  {Shibahashi}, H., {Jeffries}, S.~M., {Boumier}, P., {Gabriel}, A.~H., {Grec},
  G., {Robillot}, J.~M., Roca~Cort{\'e}s, T., \& {Ulrich}, R.~K., 1998, ApJ,
  504, L51.

\bibitem[Gelly et~al.(2002)Gelly, {Lazrek}, {Grec}, {Ayad}, {Schmider},
  {Renaud}, {Salabert}, \& {Fossat}]{GLG2002}
Gelly, B., {Lazrek}, M., {Grec}, G., {Ayad}, A., {Schmider}, F.~X., {Renaud},
  C., {Salabert}, D., \& {Fossat}, E., 2002, A\&A, 394, 285.

\bibitem[Gough(2003)]{Gou2003}
Gough, D.~O., 2003, Ap\&SS, 284, 165.

\bibitem[Harvey(1984)]{Har84}
Harvey, J.~W., 1984, In Noyes, R.~W., \& Rhodes, E.~J., editors, {\em Probing
  the depths of a star: the study of solar oscillations from space}. JPL
  400-237, NASA, JPL, Los Angeles.

\bibitem[Houdek et~al.(1999)Houdek, {Balmforth}, {Christensen-Dalsgaard}, \&
  {Gough}]{HBChD99}
Houdek, G., {Balmforth}, N.~J., {Christensen-Dalsgaard}, J., \& {Gough}, D.~O.,
  1999, A\&A, 351, 582.

\bibitem[Isaak et~al.(1989)Isaak, McLeod, Pall{\'e}, van~der Raay, \&
  Roca~Cort{\'e}s]{IMP89}
Isaak, G.~R., McLeod, C.~P., Pall{\'e}, P.~L., van~der Raay, H.~B., \&
  Roca~Cort{\'e}s, T., 1989, A\&A, 208, 297.

\bibitem[Kjeldsen et~al.(1999)Kjeldsen, Bedding, Frandsen, \& Dall]{KBF99}
Kjeldsen, H., Bedding, T.~R., Frandsen, S., \& Dall, T.~H., 1999, MNRAS, 303,
  579.

\bibitem[Kumar et~al.(1990)Kumar, {Duvall}, {Harvey}, {Jefferies}, {Pomerantz},
  \& {Thompson}]{KDH90}
Kumar, P., {Duvall}, T.~L., {Harvey}, J.~W., {Jefferies}, S.~M., {Pomerantz},
  M.~A., \& {Thompson}, M.~J., 1990, In Osaki, Y., \& Shibahashi, H., editors,
  {\em Proc.\ Oji Int.\ Seminar, Progress of Seismology of the Sun and Stars},
  Lecture Notes in Physics, Vol. 367, page~87. Berlin: Springer.

\bibitem[Kumar \& {Lu}(1991)Kumar, \& {Lu}]{K+L91}
Kumar, P., \& {Lu}, E., 1991, ApJ, 375, L35.

\bibitem[Lazrek et~al.(1997)Lazrek, {Baudin}, {Bertello}, {Boumier}, {Charra},
  {Fierry-Fraillon}, {Fossat}, {Gabriel}, {Garc{\'{\i}}a}, {Gelly}, {Gouiffes},
  {Grec}, {Palle}, {Perez Hernandez}, {Regulo}, {Renaud}, {Robillot}, {Roca
  Cortes}, {Turck-Chieze}, \& {Ulrich}]{LBB97}
Lazrek, M., {Baudin}, F., {Bertello}, L., {Boumier}, P., {Charra}, J.,
  {Fierry-Fraillon}, D., {Fossat}, E., {Gabriel}, A.~H., {Garc{\'{\i}}a},
  R.~A., {Gelly}, B., {Gouiffes}, C., {Grec}, G., {Palle}, P.~L., {Perez
  Hernandez}, F., {Regulo}, C., {Renaud}, C., {Robillot}, J.-M., {Roca Cortes},
  T., {Turck-Chieze}, S., \& {Ulrich}, R.~K., 1997, Sol. Phys., 175, 227.

\bibitem[McMillan et~al.(1993)McMillan, {Moore}, {Perry}, \& {Smith}]{McMMP93}
McMillan, R.~S., {Moore}, T.~L., {Perry}, M.~L., \& {Smith}, P.~H., 1993, ApJ,
  403, 801.

\bibitem[Meunier \& {Kosovichev}(2003)Meunier, \& {Kosovichev}]{M+K2003}
Meunier, N., \& {Kosovichev}, A., 2003, A\&A, 412, 541.

\bibitem[Ot{\'i}~Floranes et~al.(2005)Ot{\'i}~Floranes,
  {Christensen-Dalsgaard}, \& {Thompson}]{OFChDT2005}
Ot{\'i}~Floranes, H., {Christensen-Dalsgaard}, J., \& {Thompson}, M.~J., 2005,
  MNRAS, 356, 671.

\bibitem[Palle et~al.(1992)Palle, {Regulo}, {Roca-Cortes}, {Sanchez-Duarte}, \&
  {Schmider}]{PRRC92}
Palle, P.~L., {Regulo}, C., {Roca-Cortes}, T., {Sanchez-Duarte}, L., \&
  {Schmider}, F.~X., 1992, A\&A, 254, 348.

\bibitem[Stello et~al.(2004)Stello, Kjeldsen, Bedding, De~Ridder, Aerts,
  Carrier, \& Frandsen]{SKB2004}
Stello, D., Kjeldsen, H., Bedding, T.~R., De~Ridder, J., Aerts, C., Carrier,
  F., \& Frandsen, S., 2004, Sol. Phys., 220, 207.

\bibitem[Ulrich(1986)]{Ulr86}
Ulrich, R.~K., 1986, ApJ, 306, L37.

\bibitem[{Ulrich}(1988)]{Ulr88}
{Ulrich}, R.~K., 1988, In Christensen-Dalsgaard, J., \& Frandsen, S., editors,
  {\em Proc.\ IAU Symp.\ 123, Advances in Helio- and Asteroseismology}, page
  299. Dordrecht: Kluwer.

\bibitem[Ulrich et~al.(2000)Ulrich, Garc{\'{\i}}a, Robillot, {Turck-Chi{\`
  e}ze}, {Bertello}, {Charra}, {Dzitko}, {Gabriel}, \& {Roca Cort{\'
  e}s}]{UGR2000}
Ulrich, R.~K., Garc{\'{\i}}a, R.~A., Robillot, J.-M., {Turck-Chi{\` e}ze}, S.,
  {Bertello}, L., {Charra}, J., {Dzitko}, H., {Gabriel}, A.~H., \& {Roca
  Cort{\' e}s}, T., 2000, A\&A, 364, 799.

\bibitem[Vogt(1987)]{Vog87}
Vogt, S.~S., 1987, PASP, 99, 1214.

\end{thebibliography}
\end{document}